\documentclass[12pt,epsf]{article}  
\usepackage{epsf}
\usepackage{amsmath}
\usepackage{amssymb}
\usepackage{graphicx}
\begin{document}
\newcommand{\sheptitle}
{Charged lepton contributions to bimaximal and tri-bimaximal mixing for
  generating $\sin\theta_{13}\neq 0$ and $\tan^2\theta_{23}<1$}
\newcommand{\shepauthor}
{Chandan Duarah$^a$\footnote{\it{Corresponding author:} chandan001@rediffmail.com\\ 
                           \it{Postal Address:} Department of Physics, Gauhati University, Guwahati-781014, Assam, India\\
                            \it{Telephone:} $+91-0361-2570531$\\
                             \it{Fax:} $+91-0361-2700311$}, Abhijeet Das$^b$ and N. Nimai Singh$^c$}
\newcommand{\shepaddress}
   { $^{a,c}$Department of Physics, Gauhati University,
     Guwahati-781014, India \\
   $^b$Department of Physics, Assam University, Diphu Campus, Diphu-782460, Assam, India}
\newcommand{\shepabstract}
{Bimaximal(BM) and Tri-bimaximal(TB) mixings of neutrinos are two
 special cases of lepton mixing matrix, which predict the reactor angle
 $\theta_{13}=0$ and the atmospheric angle  $\tan^2 \theta_{23}=1$.
 Recent precision measurements and global analysis of
 oscillation parameters, have confirmed a non-vanishing value of
 $\theta_{13}$ as well as deviations of $\theta_{12}$ and
 $\theta_{23}$
from their maximal values predicted by BM or TB mixing. In this work
we mainly concentrate on $\theta_{13}$ and $\theta_{23}$ to assign
$\theta_{13}\neq 0$ and $\tan^2 \theta_{23}<1$ with the help of charged
lepton corrections defined by $U_{PMNS}=U_l^{\dagger} U_{\nu}$. 
We first consider $U_{\nu}$ to be given separately
by BM and TB mixing matrices and then find the possible forms of $U_l$ such
that the elements of PMNS matrix, finally yield $\theta_{13}\neq 0$ and
$\tan^2 \theta_{23}<1$ in agreement with latest observational data.
To compute the values of mixing angles we assume the charged lepton
correction to be of Cabbibo-Kobayashi-Maskawa(CKM) like. All the
mixing matrices involved in the calculation satisfy the unitarity
condition to leading order of expansion parameter.\\
Key-words: Charged lepton correction, Bimaximal and Tri-bimaximal mixing.\\
PACS number: 14.60 Pq}
\begin{titlepage}
\begin{flushright}
\end{flushright}
\begin{center}
{\large{\bf\sheptitle}}
\bigskip\\
\shepauthor
\\
\mbox{}\\
{\it\shepaddress}\\
\vspace{.5in}
{\bf Abstract}
\bigskip
\end{center}
\setcounter{page}{0}
\shepabstract
\end{titlepage}

\section{Introduction}
Recent precision measurements[1-4] and latest global $3\nu$ oscillation analysis[5] of neutrino
mixing parameters, have confirmed non-vanishing value of
$\theta_{13}$ as well as deviation of atmospheric mixing angle from
maximal value, $\theta_{23}<\pi/4$. One of the important aspects of
neutrino physics is to understand such mixing patterns[6]. Charged lepton
corrections[7] to neutrino mixing matrix is an attractive tool which can
impart non-zero value of $\theta_{13}$ as well as deviation of
$\theta_{23}$ from maximal value. We address the issue of charged
lepton correction to both bimaximal(BM) and tri-bimaximal(TB) neutrino mixings
to produce desired results.

To begin with we start with the lepton mixing matrix, known as Pontecorvo-Maki-Nakagawa-Sakata
(PMNS) matrix[8],
\begin{equation}
U_{PMNS}=U_l^{\dagger} U_{\nu},
\end{equation} 
 which is analogous to CKM matrix, $V_{CKM}=U_{uL}^{\dagger} U_{dL}$ for
 quark sector[9,10].
In relation (1), $U_{l}$ and $U_{\nu}$ are the diagonalizing matrices 
for charged lepton and left-handed Majorana neutrino mass matrices
respectively which are defined as :
 $m_l=U_{lL} m_l^{diag} V_{lR}^{\dagger}$ and $m_{\nu}=U_{\nu}^{*}
m_{\nu}^{diag} U_{\nu}^{\dagger}$. In the basis where charged lepton mass matrix is 
diagonal, $m_{\nu}$ is expressible as[11] 
\begin{equation}
m_{\nu}^{\prime}=U_{lL}^{\dagger} m_{\nu} U_{lL}.
\end{equation}
In the standard Particle Data Group (PDG) parametrization[10],
 with three mixing angles and three CP phases- one Dirac CP phase 
($\delta$) and two Majorana CP phases ($\alpha$, $\beta$), PMNS matrix has the form,

\begin{equation}
       U_{PMNS} = \begin{pmatrix}
    c_{12} c_{13}                       & s_{12} c_{13} 
                                                          & s_{13} e^{-i \delta}\\
    -s_{12} c_{23}-c_{12} s_{23} s_{13}e^{i \delta} & c_{12} c_{23}-s_{12} s_{23} s_{13} e^{i \delta}
                                                          & s_{23} c_{13}\\
    s_{12} s_{23}-c_{12} c_{23} s_{13}e^{i \delta} & -c_{12} s_{23}-s_{12} c_{23} s_{13} e^{i \delta} 
                                                          & c_{23} c_{13} \\ 
                                                      \end{pmatrix}. P,
\end{equation}
where $c_{ij}=\cos \theta_{ij}$, $s_{ij}=\sin \theta_{ij}$ with
$\theta_{12}$
 being the solar angle, $\theta_{23}$ being the atmospheric angle and
 $\theta_{13}$
 being the reactor angle and $P=diag(1, e^{i \alpha}, e^{i \beta})$
 contains the
 Majorana CP phases. In our present work we ignore all the CP phases. Then
 under
 $\mu-\tau$ symmetry, with $\theta_{13}=0$, PMNS matrix takes the form[12] :

\begin{equation}
 U_{PMNS}= \begin{pmatrix}
                    c_{12} & s_{12} & 0 \\
                    - \frac{s_{12}}{\sqrt{2}} & \frac{c_{12}}{\sqrt{2}} & \frac{1}{\sqrt{2}} \\
                     \frac{s_{12}}{\sqrt{2}} & -\frac{c_{12}}{\sqrt{2}} & \frac{1}{\sqrt{2}} \\
       \end{pmatrix},
\end{equation} 
which predicts maximal value of the atmospheric angle
 ($\theta_{23}=\displaystyle \frac{\pi}{4}$) leaving solar 
angle ($\theta_{12}$) arbitrary.\\

 Two popular neutrino 
mixing matrices are the bi-maximal(BM) mixing[13] and the
tri-bimaximal
 (TB) mixing[14], which can be obtained from equation (4) by setting
 $s_{12}=\displaystyle \frac{1}{\sqrt{2}}$ and $s_{12}=\displaystyle
\frac{1}{\sqrt{3}}$
 respectively and are given as:

\begin{equation}
U_{BM}=\begin{pmatrix}
                   \frac{1}{\sqrt{2}} & \frac{1}{\sqrt{2}} & 0 \\
                    - \frac{1}{2} & \frac{1}{2} & \frac{1}{\sqrt{2}} \\
                     \frac{1}{2} & -\frac{1}{2} & \frac{1}{\sqrt{2}} \\
       \end{pmatrix}, \  \  \ U_{TB}=\begin{pmatrix}
                                         \sqrt{\frac{2}{3}} & \frac{1}{\sqrt{3}} & 0 \\
                                          - \sqrt{\frac{1}{6}} & \frac{1}{\sqrt{3}} & \frac{1}{\sqrt{2}} \\
                                          \sqrt{\frac{1}{6}} & -\frac{1}{\sqrt{3}} & \frac{1}{\sqrt{2}} \\
                                         \end{pmatrix}.
\end{equation}
Both these two neutrino mixing matrices predict
 $\tan^2 \theta_{23}=\displaystyle \frac{|U_{\mu 3}|^2}{|U_{\tau
    3}|^2}=1$
 and $\sin^2 \theta_{13}=|U_{e3}|^2=0$ .

 The paper is organized as follows: In section 2 we discuss charged
 lepton
 correction to BM neutrino mixing and present predictions of the
 mixing
 angles along with graphical representations.
 In a similar way section 3 is devoted to TB mixing. Then in section 4
 we analyze both the schemes in presence of Dirac CP phase. Finally
 section 4 is devoted to summary and discussion.


\section{Charged lepton correction to BM mixing}

General forms of the lepton mixing matrix ($U_{\nu}$)
 and the neutrino mixing matrix ($U_l$) in equation(1) can be expressed as 
\begin{equation}
 U_l = \begin{pmatrix}
    c_{12}^l c_{13}^l  & s_{12}^l c_{13}^l & s_{13}^l \\
 -s_{12}^l c_{23}^l-c_{12}^l s_{23}^l s_{13}^l & c_{12}^l
 c_{23}^l-s_{12}^l s_{23}^l s_{13}^l & s_{23}^l c_{13}^l\\
 s_{12}^l s_{23}^l-c_{12}^l c_{23}^l s_{13}^l & -c_{12}^l s_{23}^l-s_{12}^l
 c_{23}^l s_{13}^l & c_{23}^l c_{13}^l \\ 
 \end{pmatrix}
\end{equation}
and
\begin{equation}
 U_{\nu} = \begin{pmatrix}
    c_{12}^{\nu} c_{13}^{\nu} & s_{12}^{\nu} c_{13}^{\nu} & s_{13}^{\nu} \\
     -s_{12}^{\nu} c_{23}^{\nu}-c_{12}^{\nu} s_{23}^{\nu} s_{13}^{\nu}
     &
 c_{12}^{\nu} c_{23}^{\nu}-s_{12}^{\nu} s_{23}^{\nu} s_{13}^{\nu} & 
s_{23}^{\nu} c_{13}^{\nu}\\
    s_{12}^{\nu} s_{23}^{\nu}-c_{12}^{\nu} c_{23}^{\nu} s_{13}^{\nu} &
 -c_{12}^{\nu} s_{23}^{\nu}-s_{12}^{\nu} c_{23}^{\nu} s_{13}^{\nu} &
 c_{23}^{\nu} c_{13}^{\nu} \\ 
                                                      \end{pmatrix} ,
\end{equation} 
where we have ignored the CP violating phases. For our case
 we first consider the neutrino mixing pattern to be of bi-maximal 
nature. Then $U_{\nu}=U_{BM}$ is given by equation (5).
We then take the following form of the lepton mixing matrix[15],
\begin{equation}
U_{l}=\begin{pmatrix}
                   \tilde{c}_{12} & \tilde{s}_{12} & 0 \\
                 - \tilde{s}_{12} & \tilde{c}_{12} & 0 \\
                     0               &             0 & 1 \\
       \end{pmatrix},
\end{equation}
where $\tilde{s}_{ij}=\sin \theta^l_{ij}$ and $\tilde{c}_{ij}=\cos
\theta^l_{ij}$. This structure(8) had been studied earlier[15] but we
study  it again here in the light of latest observational data[5].

\begin{table}[tbp]
\begin{center}
\begin{tabular}{|c|c|c|c|}
\hline  
parameter & best fit & $1\sigma$ range & $3\sigma$ range \\ \hline
$\tan^2 \theta_{12}$ & 0.470 & 0.435-0.506 & 0.370-0.587  \\
$\tan^2 \theta_{23}$ & 0.745 & 0.667-0.855 & 0.563-2.125 \\
$\sin^2 \theta_{13}$ & 0.0246 & 0.0218-0.0275 & 0.017-0.033 \\
 \hline
\end{tabular}
\caption{Best fit, $1\sigma$ and $3\sigma$ ranges of parameters for NH obtained
         from global analysis[22]}
\end{center}
\end{table}

From equations (1), (5) and (8), we finally obtain the PMNS matrix 
$U_{PMNS}=U_l^{\dagger} U_{BM}$ as
\begin{equation}
U_{PMNS}=\begin{pmatrix}
 \frac{1}{\sqrt{2}}(\tilde{c}_{12}+\frac{\tilde{s}_{12}}{\sqrt{2}})
                & \frac{1}{\sqrt{2}}(\tilde{c}_{12}-\frac{\tilde{s}_{12}}{\sqrt{2}}) 
                                              & -\frac{\tilde{s}_{12}}{\sqrt{2}} \\
   - \frac{1}{2}(\tilde{c}_{12}-\sqrt{2}\tilde{s}_{12}) 
                & \frac{1}{2}(\tilde{c}_{12}+\sqrt{2}\tilde{s}_{12})
                                      & \frac{\tilde{c}_{12}}{\sqrt{2}} \\
                    \frac{1}{2} & -\frac{1}{2} & \frac{1}{\sqrt{2}} \\
       \end{pmatrix}.
\end{equation}
Let us now assume that the charged lepton corrections are
 Cabbibo-Kobayashi-Maskawa (CKM) like[10], which allows us to take
\begin{equation}
\tilde{s}_{12}=\sin \theta^l_{12}=\lambda,
\end{equation}
where the Wolfestein parameter $\lambda$ is related to the Cabbibo angle ($\theta_C$) by
 $\lambda=\sin \theta_C$. Under this consideration, PMNS matrix in
equation (9), can be approximated to the form,
\begin{equation}
U_{PMNS} \approx \begin{pmatrix}
       \frac{1}{\sqrt{2}}(1+\frac{\lambda}{\sqrt{2}}-\frac{\lambda^2}{2})
                    & \frac{1}{\sqrt{2}}(1-\frac{\lambda}{\sqrt{2}}-\frac{\lambda^2}{2})
                                 & -\frac{\lambda}{\sqrt{2}} \\
         - \frac{1}{2}(1-\sqrt{2}\lambda-\frac{\lambda^2}{2})
                    & \frac{1}{2}(1+\sqrt{2}\lambda-\frac{\lambda^2}{2})
                                 & \frac{1}{\sqrt{2}}(1-\frac{\lambda^2}{2}) \\
                    \frac{1}{2} & -\frac{1}{2} & \frac{1}{\sqrt{2}} \\
       \end{pmatrix}.
\end{equation}

\begin{figure}
\begin{center}
\includegraphics[scale=1]{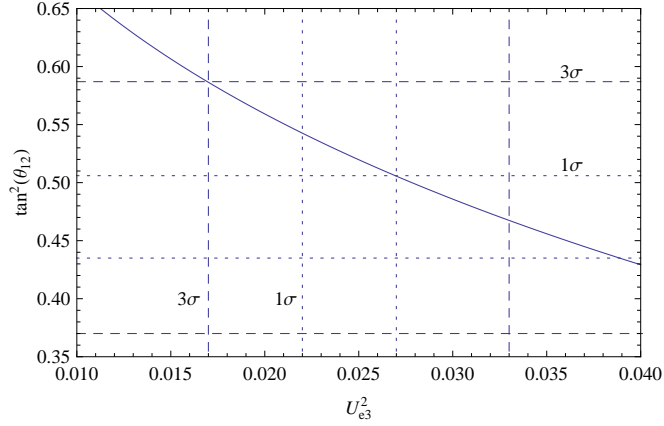}
\caption{Variation of $\tan^2 \theta_{12}$ with $U_{e3}^2$ for BM
  mixing after taking charged lepton correction. Dotted and Dashed lines
  represents $1\sigma$ and $3\sigma$ bounds respectively, 
  obtained from the global analysis[22]}
\end{center}
\end{figure}

And the expression in equation (8) becomes 
\begin{equation}
U_{lL}= \begin{pmatrix}
           1-\frac{\lambda^2}{2} & -\lambda              & 0 \\
            \lambda              & 1-\frac{\lambda^2}{2} & 0 \\
            0                    &             0         & 1 \\
      \end{pmatrix} .
\end{equation}
It can be emphasised here that both mixing matrices in equations (11) and
(12) satisfy the unitarity condition as expected.
Then equation (11) leads to 
\begin{equation}
\tan^2\theta_{12} = \left( \frac{1-|U_{e3}|-|U_{e3}|^2}{1+|U_{e3}|-|U_{e3}|^2}\right)^2 ,
\end{equation}

\begin{equation}
\tan^2 \theta_{23} = ( 1-|U_{e3}|^2 )^2 ,
\end{equation}

\begin{equation}
|U_{e3}|^2 = \sin^2 \theta_{13}=\frac{\lambda^2}{2}.
\end{equation}
With $\lambda=0.232$ corresponding to $|U_{e3}|^2=0.027$,
 we get $\tan^2 \theta_{12} \approx 0.50$ and $\tan^2 \theta_{23}
 =0.946$. The variations of $\tan^2 \theta_{12} $ with $|U_{e3}|^2$ and
 $\tan^2 \theta_{23} $ with $|U_{e3}|^2$ are shown in Fig.1 and
Fig.2 respectively for both $1\sigma$ and $3\sigma$ ranges (Table 1) of latest
global observational data [22]. As expected $3\sigma$ range of data
can accomodate both $\tan^2 \theta_{12} $ and $\tan^2 \theta_{23}$
predictions. However, the $1\sigma$ range of data just marginally
covers $\tan^2 \theta_{12}$ prediction at $\tan^2 \theta_{12} \approx 0.5$
(TB value) but not the $\tan^2 \theta_{23} $ prediction within the range.
Certain theoretical refinements are needed in this front.

\begin{figure}
\begin{center}
\includegraphics[scale=1]{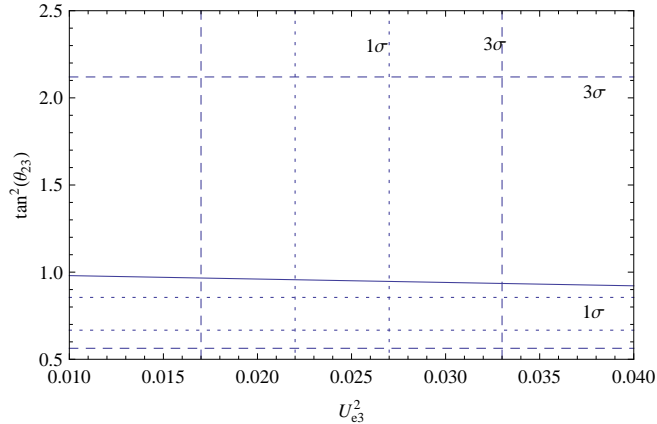}
\caption{Variation of $\tan^2 \theta_{23}$ with $U_{e3}^2$ for BM
  mixing after taking charged lepton correction. Dotted and Dashed lines
  represents $1\sigma$ and $3\sigma$ bounds respectively, 
  obtained from the global analysis [22]  }
\end{center}
\end{figure}

\section{Charged lepton correction to TB Mixing}

Tri-bimaximal neutrino mixing is a special case of mixing matrix 
with $\mu-\tau$ symmetry. It can give a very close description 
of the experimental data except the case: $\theta_{13}=0$. 
The TB neutrino mixing matrix ($U_{\nu}=U_{TB}$) is given in equation
(5). In order to account for the charged lepton correction to the TB neutrino 
mixing, we start with the lepton mixing matrix which satisfies
unitarity condition, 

\begin{equation}
\tilde{U}_l=\begin{pmatrix}
1-\frac{\lambda^2}{4} & -\frac{\lambda}{2} & -\frac{\lambda}{2} \\
 \frac{\lambda}{2} & 1-\frac{\lambda^2}{8} & -\frac{\lambda^2}{8} \\
 \frac{\lambda}{2} & -\frac{\lambda^2}{8} & 1-\frac{\lambda^2}{8} \\
 \end{pmatrix}.
\end{equation}
Using the form of $U_{\nu}$ for TB, given by equation (5), we have
 $U_{PMNS}=\tilde{U}^{\dagger}_l U_{TB}$ which  reproduces the following PMNS
matrix first proposed by King[16],
\begin{equation}
U_{PMNS}=\begin{pmatrix}
      \sqrt{\frac{2}{3}}(1-\frac{\lambda^2}{4}) &
      \frac{1}{\sqrt{3}}(1-\frac{\lambda^2}{4}) &
 \frac{\lambda}{\sqrt{2}} \\
   -\frac{1}{\sqrt{6}}(1+\lambda) &
   \frac{1}{\sqrt{3}}(1-\frac{\lambda}{2}) &
  \frac{1}{\sqrt{2}}(1-\frac{\lambda^2}{4}) \\
     \frac{1}{\sqrt{6}}(1-\lambda) &
     -\frac{1}{\sqrt{3}}(1+\frac{\lambda}{2}) & 
\frac{1}{\sqrt{2}}(1-\frac{\lambda^2}{4})\\
                                         \end{pmatrix}.
\end{equation}
This PMNS matrix has unique property of unitarity to leading order,
and also predicts $\tan^2 \theta_{23}=1$. In order to have
$\tan^2 \theta_{23}<1$ in the light of present experimental data[5],
 we now modify the charged lepton mixing matrix(16) by the relation

\begin{equation}
U_l^{\dagger}=\tilde{R}_{23}^{\dagger} \tilde{U}_l^{\dagger},
\end{equation}
where $\tilde{R}_{23}$ has a structure similar to that of rotation matrix and is given by

\begin{equation}
\tilde{R}_{23}= \begin{pmatrix}
           1  &  0   &  0 \\
           0  & \tilde{c}_{23} & \tilde{s}_{23}\\
           0  & -\tilde{s}_{23} & \tilde{c}_{23}\\
\end{pmatrix},
\end{equation}
with $\tilde{s}_{23}=\sin \theta^l_{23}$ and $\tilde{c}_{23}=\cos
\theta^l_{23} $. 

\begin{figure}
\begin{center}
\includegraphics[scale=1]{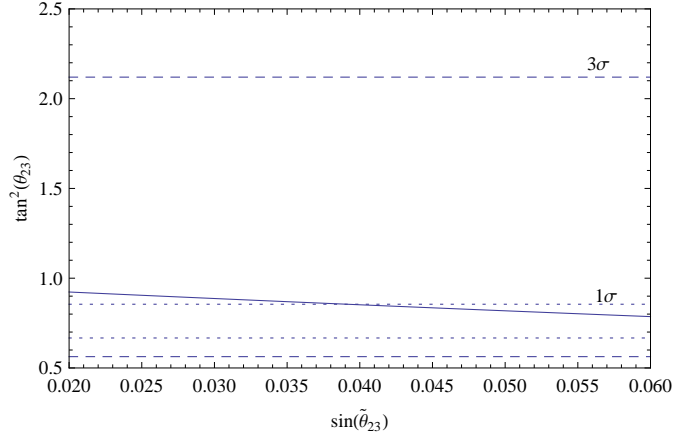}
\caption{Variation of $\tan^2 \theta_{23}$ with $\sin \tilde{\theta}_{23}$
 for TB mixing after taking charged lepton correction. Dotted and Dashed lines
  represents $1\sigma$ and $3\sigma$ bounds respectively, 
  obtained from the global analysis[22] }
\end{center}
\end{figure}

\begin{figure}
\begin{center}
\includegraphics[scale=1]{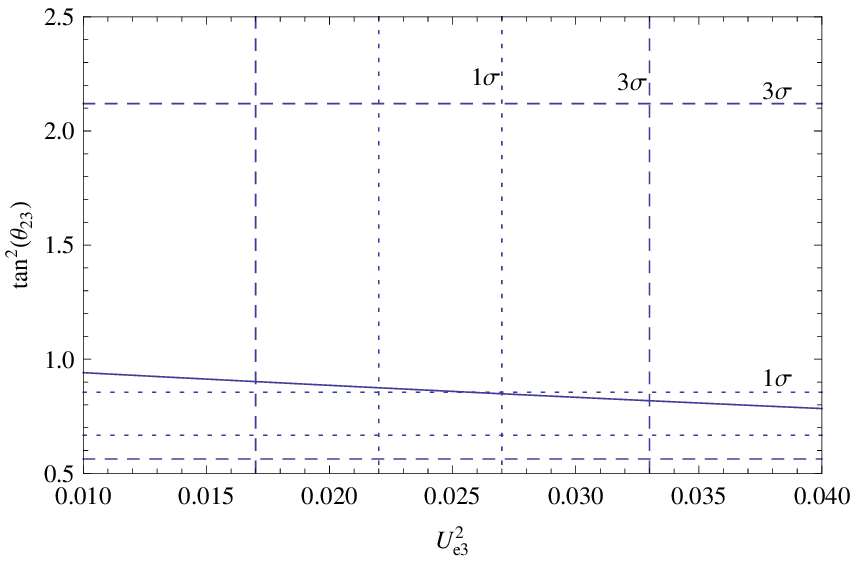}
\caption{Variation of $\tan^2 \theta_{23}$ with $U_{e3}^2$  for TB
  mixing after taking charged lepton correction. Dotted and Dashed lines
  represents $1\sigma$ and $3\sigma$ bounds respectively, 
  obtained from the global analysis[22] }
\end{center}
\end{figure}
Then equations (1),(16) and (18) give the following elements of the new
PMNS matrix, $U_{PMNS}={U}^{\dagger}_l U_{TB}$,

\begin{align*}
 &(U_{PMNS})_{11}= \sqrt{\frac{2}{3}}(1-\frac{\lambda^2}{4}),\\
    &(U_{PMNS})_{12}= \frac{1}{\sqrt{3}}(1-\frac{\lambda^2}{4}),\\
    &(U_{PMNS})_{13}= \frac{\lambda}{\sqrt{2}}, \\
    &(U_{PMNS})_{21}=-\frac{1}{\sqrt{6}}[(\tilde{c}_{23}+\tilde{s}_{23})+(\tilde{c}_{23}-\tilde{s}_{23})\lambda],\\
   &(U_{PMNS})_{22}= \frac{1}{\sqrt{3}}[(\tilde{c}_{23}+\tilde{s}_{23})-(\tilde{c}_{23}-\tilde{s}_{23})\frac{\lambda}{2}],\\
    &(U_{PMNS})_{23}= \frac{1}{\sqrt{2}}(\tilde{c}_{23}-\tilde{s}_{23})(1-\frac{\lambda^2}{4}), \\
    &(U_{PMNS})_{31}= \frac{1}{\sqrt{6}}[(\tilde{c}_{23}-\tilde{s}_{23})-(\tilde{c}_{23}+\tilde{s}_{23})\lambda], \\
 &(U_{PMNS})_{32}= -\frac{1}{\sqrt{3}}[(\tilde{c}_{23}-\tilde{s}_{23})+(\tilde{c}_{23}+\tilde{s}_{23})\frac{\lambda}{2}],\\
    &(U_{PMNS})_{33}= \frac{1}{\sqrt{2}}(\tilde{c}_{23}+\tilde{s}_{23})
                   (1-\frac{\lambda^2}{4}).\ \ \ \ \ \ \ \ \
  \ \ \ \ \ \ \ \ \ \ \ \ \ \ \ \ \ \ \ \ \ \ \ \ \ \ \ \  (A) 
\end{align*}

From these elements we calculate 
\begin{equation}
\tan^2 \theta_{23}=\displaystyle \left( \frac{1-\tan \tilde{\theta}_{23}}{1+\tan \tilde{\theta}_{23}}\right)^2,
\end{equation}
which is lesser than maximal value for non-zero $\tan \tilde{\theta}_{23}$.
Assuming that the charged lepton corrections are
Cabbibo-Kobayashi-Maskawa (CKM) like,  we can have[10,17] 

\begin{equation}
\tilde{s}_{23}=\sin \theta^l_{23}=A \lambda^2 \approx 0.041,
\end{equation}
leading to $\tan^2 \theta_{23}=0.85$, where we have adopted
$\lambda=0.2324$ and $A=0.759$.
 The variation of $\tan^2 \theta_{23}$ with $\sin
 \tilde{\theta}_{23}$ is shown in Fig.3. The prediction on $\tan^2 \theta_{12}$
is fixed at TB value while the change is confined to $\tan^2 \theta_{23}$
only and its variation with $|U_{e3}|^2$ along with $1\sigma$ and $3\sigma$ ranges 
of latest global observational data[22] is shown in Fig.4. At $3\sigma$
range the prediction on $\tan^2 \theta_{23}$ is in fair agreement with 
global data as like BM case. However, in TB case we notice an improvement of our prediction 
at $1\sigma$ range that it just passes through the $1\sigma$ region in the plot 
unlike the BM case.

\section{Effects of Dirac CP phase}

In this section we would like to discuss briefly the effects of CP 
violating phases in the proposed schemes.
To observe the effects of the Dirac type CP phase in the BM scheme we follow
 two ways of introducing the phase. First case assumes a CP phase $\phi$, 
 coming from the charged lepton sector, with the unitary matrix [20]
\begin{equation}
U_{l}=\begin{pmatrix}
                   \tilde{c}_{12}            & \tilde{s}_{12} e^{-i \phi} & 0 \\
                 - \tilde{s}_{12} e^{i \phi} & \tilde{c}_{12} & 0 \\
                     0               &             0 & 1 \\
       \end{pmatrix}.
\end{equation}
With this $U_l$, $U_{PMNS}=U_l^{\dagger} U_{BM}$ yields
\begin{equation}
U_{PMNS}=\begin{pmatrix}
 \frac{1}{\sqrt{2}}(\tilde{c}_{12}+\frac{\tilde{s}_{12}}{\sqrt{2}} e^{i \phi}) 
             & \frac{1}{\sqrt{2}}(\tilde{c}_{12}-\frac{\tilde{s}_{12}}{\sqrt{2}} e^{i \phi}) 
                                          & -\frac{\tilde{s}_{12}}{\sqrt{2}} e^{i \phi} \\
   - \frac{1}{2}(\tilde{c}_{12}-\sqrt{2}\tilde{s}_{12}e^{-i \phi}) 
                & \frac{1}{2}(\tilde{c}_{12}+\sqrt{2}\tilde{s}_{12}e^{-i \phi})
                                      & \frac{\tilde{c}_{12}}{\sqrt{2}} \\
                    \frac{1}{2} & -\frac{1}{2} & \frac{1}{\sqrt{2}} \\
       \end{pmatrix}.
\end{equation}
In the second approach we introduce the CP phase $\delta$, originating from neutrino 
sector, by the following relation [21],
\begin{equation}
U_{PMNS}=U_l^{\dagger}R_{23}Diag(e^{i \delta},1,e^{-i \delta})R_{12},
\end{equation}
where $U_l$ is given by equation (8) and $R_{23}$ and $R_{12}$ are the $3 \times 3$ 
orthogonal rotation matrices with $\displaystyle \theta_{23}=\frac{\pi}{4} $ and
$\displaystyle \theta_{12}=\frac{\pi}{4} $ respectively. Then equation (24) gives,
 \begin{equation}
U_{PMNS}=\begin{pmatrix}
 \frac{1}{\sqrt{2}}(\tilde{c}_{12}e^{i \delta}+\frac{\tilde{s}_{12}}{\sqrt{2}})
             & \frac{1}{\sqrt{2}}(\tilde{c}_{12}e^{i \delta}-\frac{\tilde{s}_{12}}{\sqrt{2}}) 
                                      & -\frac{\tilde{s_{12}}}{\sqrt{2}}e^{-i \delta} \\
   - \frac{1}{2}(\tilde{c}_{12}-\sqrt{2}\tilde{s}_{12}e^{i \delta}) 
              & \frac{1}{2}(\tilde{c}_{12}+\sqrt{2}\tilde{s}_{12}e^{i \delta})
                                & \frac{\tilde{c}_{12}}{\sqrt{2}}e^{-i \delta} \\
                    \frac{1}{2} & -\frac{1}{2} & \frac{1}{\sqrt{2}}e^{-i \delta} \\
       \end{pmatrix}.
\end{equation}
Both the cases lead to a similar form of the rephasing invariant quantity defined as
$J_{CP}=Im\{U_{e2}U_{\mu 3}U^{*}_{e3}U^{*}_{\mu 2}\}$. For example, we get

\begin{equation}
J^{BM}_{CP}=\displaystyle \frac{1}{4 \sqrt{2}} \sin \tilde{\theta}_{12}
                    \cos \tilde{\theta}_{12} \sin \phi,
\end{equation}
and
\begin{equation}
J^{BM}_{CP}=\displaystyle \frac{1}{4 \sqrt{2}} \sin \tilde{\theta}_{12}
                    \cos \tilde{\theta}_{12} \sin \delta,
\end{equation}
from equations (23) and (25) respectively. We further calculate

\begin{equation}
\tan^2 \theta_{12}=\displaystyle \frac{2-\tilde{s}^2_{12}-
                          2\sqrt{2}\tilde{c}_{12}\tilde{s}_{12}\cos \phi}
                          {2-\tilde{s}^2_{12}+
                          2\sqrt{2}\tilde{c}_{12}\tilde{s}_{12}\cos \phi}
\end{equation}
and 
\begin{equation}
\tan^2 \theta_{12}=\displaystyle \frac{2-\tilde{s}^2_{12}-
                          2\sqrt{2}\tilde{c}_{12}\tilde{s}_{12}\cos \delta}
                          {2-\tilde{s}^2_{12}+
                          2\sqrt{2}\tilde{c}_{12}\tilde{s}_{12}\cos \delta}
\end{equation}
from equations (23) and (25) respectively, which show the dependence of solar angle 
on the CP phase. For maximal CP violation ($\sin \delta=\pm 1$) we get 
$|J^{BM}_{CP}|_{max} \approx 0.03989$. From the relation $\sin \tilde{\theta}_{12}
=\sqrt{2} \sin \theta_{13}$ along with the approximation $\cos \tilde{\theta}_{12} 
\approx 1$ (from eq.(10)) equation (27) gives 
\begin{equation}
J^{BM}_{CP} \approx \displaystyle \frac{1}{4} \sin \theta_{13} \sin \delta
\end{equation}
which is consistent with the result of reference[7].\\

To incorporate the Dirac type CP effects in TB scheme we first adopt the Tri-bimaximal-Cabbibo 
mixing matrix $U_{TBC}$ proposed by King[16].

\begin{equation}
U_{TBC}=\begin{pmatrix}
      \sqrt{\frac{2}{3}}(1-\frac{\lambda^2}{4})
             & \frac{1}{\sqrt{3}}(1-\frac{\lambda^2}{4}) 
                       & \frac{\lambda}{\sqrt{2}}e^{-i \delta} \\
   -\frac{1}{\sqrt{6}}(1+\lambda e^{i \delta})
              & \frac{1}{\sqrt{3}}(1-\frac{\lambda}{2}e^{i \delta}) 
                      &  \frac{1}{\sqrt{2}}(1-\frac{\lambda^2}{4}) \\
     \frac{1}{\sqrt{6}}(1-\lambda e^{i \delta}) 
              & -\frac{1}{\sqrt{3}}(1+\frac{\lambda}{2}e^{i \delta}) 
                        & \frac{1}{\sqrt{2}}(1-\frac{\lambda^2}{4})\\
                                         \end{pmatrix}.
\end{equation}
For $\delta=0$ equation (30) reproduces the mixing matrix given by equation (17). 
Then the relation $U_{PMNS}=\tilde{R}^{\dagger}_{23} U_{TBC}$ produces the
following desired elements of the PMNS matrix, given in the set of equations (A), modified 
by the CP phase $\delta$.

\begin{align*}
 &(U_{PMNS})_{11}= \sqrt{\frac{2}{3}}(1-\frac{\lambda^2}{4}),\\
    &(U_{PMNS})_{12}= \frac{1}{\sqrt{3}}(1-\frac{\lambda^2}{4}),\\
    &(U_{PMNS})_{13}= \frac{\lambda}{\sqrt{2}}e^{-i \delta}, \\
    &(U_{PMNS})_{21}=-\frac{1}{\sqrt{6}}[(\tilde{c}_{23}+\tilde{s}_{23})+(\tilde{c}_{23}-\tilde{s}_{23})\lambda e^{i \delta} ],\\
   &(U_{PMNS})_{22}= \frac{1}{\sqrt{3}}[(\tilde{c}_{23}+\tilde{s}_{23})-(\tilde{c}_{23}-\tilde{s}_{23})\frac{\lambda}{2}e^{i \delta}],\\
    &(U_{PMNS})_{23}= \frac{1}{\sqrt{2}}(\tilde{c}_{23}-\tilde{s}_{23})(1-\frac{\lambda^2}{4}), \\
    &(U_{PMNS})_{31}= \frac{1}{\sqrt{6}}[(\tilde{c}_{23}-\tilde{s}_{23})-(\tilde{c}_{23}+\tilde{s}_{23})\lambda e^{i \delta}], \\
 &(U_{PMNS})_{32}= -\frac{1}{\sqrt{3}}[(\tilde{c}_{23}-\tilde{s}_{23})+(\tilde{c}_{23}+\tilde{s}_{23})\frac{\lambda}{2}e^{i \delta}],\\
    &(U_{PMNS})_{33}= \frac{1}{\sqrt{2}}(\tilde{c}_{23}+\tilde{s}_{23})
                                       (1-\frac{\lambda^2}{4}). \ \ \ \ \ \ \ \
  \ \ \ \ \ \ \ \ \ \ \ \ \ \ \ \ \ \ \ \ \ \ \ \  (B) 
\end{align*}

The set of equations (B) predicts the rephasing invariant quantity as
\begin{equation}
J^{TB}_{CP}=\displaystyle \frac{1}{6}\lambda \left(1-\frac{\lambda^2}{4}\right)^2 
                        (\tilde{c}^2_{23}-\tilde{s}^2_{23}) \sin \delta.
\end{equation}

We also examine the structure of the PMNS matrix under the parameterization
described in equation (24) where $U^{\dagger}_l$ is now given by equation (18) and 
$R_{23}$ and $R_{12}$ are respectively decribed by $\displaystyle \theta_{23}=\frac{\pi}{4}$ 
and $\displaystyle \theta_{12}=\arcsin \frac{1}{\sqrt{3}}$. We then obtain the following
elements of the PMNS matrix:

\begin{align*}
 &(U_{PMNS})_{11}= \sqrt{\frac{2}{3}}(1-\frac{\lambda^2}{4})e^{i \delta},\\
    &(U_{PMNS})_{12}= \frac{1}{\sqrt{3}}(1-\frac{\lambda^2}{4})e^{i \delta},\\
    &(U_{PMNS})_{13}= \frac{\lambda}{\sqrt{2}}e^{-i \delta}, \\
    &(U_{PMNS})_{21}=-\frac{1}{\sqrt{6}}[(\tilde{c}_{23}+\tilde{s}_{23})+(\tilde{c}_{23}-\tilde{s}_{23})\lambda e^{i \delta}],\\
   &(U_{PMNS})_{22}= \frac{1}{\sqrt{3}}[(\tilde{c}_{23}+\tilde{s}_{23})-(\tilde{c}_{23}-\tilde{s}_{23})\frac{\lambda}{2}e^{i \delta}],\\
    &(U_{PMNS})_{23}= \frac{1}{\sqrt{2}}(\tilde{c}_{23}-\tilde{s}_{23})
                                  (1-\frac{\lambda^2}{4})e^{-i \delta}, \\
    &(U_{PMNS})_{31}= \frac{1}{\sqrt{6}}[(\tilde{c}_{23}-\tilde{s}_{23})-(\tilde{c}_{23}+\tilde{s}_{23})\lambda e^{i \delta}], \\
 &(U_{PMNS})_{32}= -\frac{1}{\sqrt{3}}[(\tilde{c}_{23}-\tilde{s}_{23})+(\tilde{c}_{23}+\tilde{s}_{23})\frac{\lambda}{2}e^{i \delta}],\\
    &(U_{PMNS})_{33}= \frac{1}{\sqrt{2}}(\tilde{c}_{23}+\tilde{s}_{23})
                         (1-\frac{\lambda^{2}}{4})e^{-i \delta}. \ \ \ \ \ \ \
   \ \ \ \ \ \ \ \ \ \ \ \ \ \ \ \ \ \ \ \ \  (C) 
\end{align*}

The set of equations (C) yields the same rephasing invariant quantity in equation (32).
From the relation $\lambda=\sqrt{2} \sin \theta_{13}$ along with equations (10) and (21),
equation (32) gives
\begin{equation}
J^{TB}_{CP} \approx \displaystyle \frac{1}{3 \sqrt{2}}\sin \theta_{13} \sin \delta.
\end{equation}
 Further, for maximal CP violation, we calculate $|J^{TB}_{CP}|_{max} \approx 0.0374$ 
from equation (32). The expression for $J_{CP}$
in equation (33) is consistent with the result of reference[7].

\section{Summary and Discussion}

We have studied two possible forms of the lepton mixing matrix $U_l$
 which can produce desired deviations from the bimaximal (BM) and tri-bimaximal(TB) mixings of 
neutrino sector under charged lepton corrections. The lepton
 mixing matrices have basically been derived from rotation matrices
 and hence the conditions of unitarity of all diagonalising matrices
 including the final form of PMNS matrices discussed here, are
 satisfied at leading order. In such situation PMNS matrix proposed by King[16]
 is a pointer to the right direction.
 Asuming the charged lepton correction is CKM-like and taking
$\lambda=0.232$ we get $\sin^2\theta_{13}=0.027$ for both BM and TB cases.
For the same value of $\lambda$ we calculate
$\tan^2 \theta_{12} \approx 0.50$ and $\tan^2 \theta_{23}=0.946<1$ 
for BM case. After the introduction of Dirac CP phase 
we observe that $\tan^2 \theta_{12}$ is affected by the phase, but 
not $\tan^2 \theta_{23}$. We also find that predictions on $\tan^2 \theta_{12}$
and $\tan^2 \theta_{23}$ in terms of $|U_{e3}|^2$ are consistent 
with the $3\sigma$ range of latest global observational data. However, 
at $1\sigma$ range the predictions are not comfortable. In case of 
TB mixing, the charged lepton correction only deviates the atmospheric angle. 
The solar angle remains fixed at its TB value ($\tan^2 \theta_{23}=0.5$). 
For $\lambda=0.232$ and $A=0.759$ we get $\tan^2 \theta_{23}=0.85<1$. 
The variation of $\tan^2 \theta_{23}$ with $|U_{e3}|^2$ shows 
that at $3\sigma$ range the prediction on $\tan^2 \theta_{23}$ is smoothly
consistent with global data. However, in TB case we get better agreement
of our prediction with $1\sigma$ range of global data
than that in BM case. Unlike the BM case, the inclusion of 
Dirac CP phase in TB mixing does not affect $\tan^2 \theta_{12}$ and 
$\tan^2 \theta_{23}$. Finally we obtain two important expressions for the
rephasing invariant quantity: $J^{BM}_{CP} \approx \displaystyle \frac{1}{4} 
\sin \theta_{13} \sin \delta$ and $J^{TB}_{CP} \approx \displaystyle 
\frac{1}{3 \sqrt{2}}\sin \theta_{13} \sin \delta$ which are consistent
with the results of reference[7]. 

The  deviation of solar mixing angle $\tan^2\theta_{12}$
below the value of  $0.50$, can be introduced in realistic $\mu-\tau$
symmetric neutrino mass matrices with specific choices of value of
flavour twister term[15,18,19] present in the texture of the mass matrices, without affecting the good
predictions on reactor and atmospheric mixing angles.


\begin{thebibliography}{23}
\bibitem{1} F. P. An {\it et al.} [DAYA-BAY Collaboration],
  Phys. Rev. Lett. 108, 171803 (2012) [arXiv:1203.1669 [hep-ex]].
\bibitem{2} J. K. Ahn {\it et al.} [RENO Collaboration], Phys. Rev.
 Lett. 108, 191802 (2012) [arXiv:1204.062 [hep-ex].]
\bibitem{3} Y. Abe {\it et al.} [DOUBLE-CHOOZ Collaboration],
  Phys. Rev. Lett. 108, 131801 (2012) [arXiv:1112.6353 [hep-ex]].
\bibitem{4}The status of neutrino oscillations has been recently
reviewed in several presentations at Neutrino-2012, the XXV
Intl. Conf.on Neutrino physics and Astrophysics (Kyoto, Japan,2012),
available at the website:neu2012.kek.jp.
\bibitem{5}G. L. Fogli, E. Lisi, Marrone, D. Montanino, A. Palazzo,
  A.M. Rotunno, arXiv:1205.5254v3[hep-ph].
\bibitem{6}Guido Altarelli, Ferruccio Feruglio, Isabella Masina, Luca
  Merlo, arXiv:1207.0587v1[hep-ph].
\bibitem{7} David Marzocca, S. T. Petcov, Andrea Romanino, Martin Spinrath, JHEP11(2011)009, arXiv:1108.0614v2[hep-ph]; 
   Guido Altarelli, Ferruccio Feruglio, Luca Merlo, 0903.1940v2[hep-ph];
   Reinier de Adelhart Toorop, Federica Bazzocchi, Luca Merlo, arXiv:1003.4502v2[hep-ph];
   Guido Altarelli, Ferruccio Feruglio, Luca Merlo and Emmanuel Stamou, arXiv:1205.4670v1[hep-ph];
   Federica Bazzocchi and Luca Merlo, arXiv:1205.5135v1[hep-ph];
   S. Morisi, Ketan M. Patel, E. Peinado, Phys. Rev. D 84,053002,2011, arXiv:1107.0696v2[hep-ph];
   L. Dorame, S. Morisi, E. Peinado, J. W. F. Valle, arXiv:1203.0155v1[hep-ph];
   S. Antusch, S. F. King, arXiv:hep-ph/0508044v2; 
   Stefan Antusch, Vinzenz Maurer, Phys.Rev.D 84, 117301(2011), arXiv:1107.3728v2[hep-ph];
   Stefan Antusch, Christian Gross, Vinzenz Maurer, Constantin Sluka, arXiv:1205.1051v2[hep-ph].
\bibitem{8}B.Pontecorvo, Sov.Phys. JETP 26, 984(1968); Z. Maki,
  M. Nakagawa and S. Sakata, Prog. Theor. Phys. 28, 870 (1962).
\bibitem{9}N. Cabibbo, Phys. Rev. Lett. 10, 531 (1963),
 M. Kobayashi and T. Maskawa, Prog. Theor. Phys. 49, 652 (1973).
\bibitem{10} K. Nakamura {\it et al.} Particle Data Group, J. Phys. G:
   Nucl. Partl. Phys. 37, N0.7A (2010).
\bibitem{11} S. F. King and N. Nimai Singh, Nucl. Phys. B591, 3 (2000);
  Nucl. Phys. B596, 81 (2001).
\bibitem{12} P. F. Harrison, W. G. Scott, Phys. Lett. B547,219 (2002),
  [hep-ph/0210197].
\bibitem{13} G. Altarelli, F. Feruglio, Phys. Rep. 320, 295 (1999);
  V. D. Barger, S. Pakvasa, T. J. Weiler, K. Whisnant,
  Phys. Lett. B437, 107 (1998); N. Nimai Singh and M. Patgiri,
  Intl. J. Mod. Phys. A17, 3629 (2002).
\bibitem{14} P. F. Harrison, D. H. Perkins and W. G. Scott
  Phys. Lett. B 530, 167 (2002) [arXiv:hep-ph/0202074].
\bibitem{15} Mahadev Patgiri and N. Nimai Singh Phys. Lett. B 567, 69 (2003).
\bibitem{16} S. F. King [arXiv:1205.0506v2[hep-ph]].
\bibitem{17} Bo-Qiang Ma, [arXiv:1205.0766v2 [hep-ph]].
\bibitem{18} N. Nimai Singh, M. Rajkhowa, A. Borah, J. Phys. G:
  Nucl. Partl. Phys. 34, 345 (2007); Pramana J. Phys. 69, 533 (2007).
\bibitem{19} Ng. K. Francis and N. Nimai Singh, Nucl. Phys. B863, 19 (2012).
\bibitem{20} Zhi-zhong Xing, arXiv:hep-ph/0204049v1; Kim Siyeon, arXiv:1208.2645v1[hep-ph].
\bibitem{21} Harald Fritzsch, Zhi-zhong Xing, arXiv:hep-ph/9708366v1.
\bibitem{22} D. V. Forero, M. Tortola and J. W. F. Valle, arXiv:1205.4018v3[hep-ph] 
\end{thebibliography}
\end{document}